\documentclass[12pt]{article}
\usepackage{amssymb}
\usepackage{setspace}
\newcommand{\nc}{\newcommand}
\nc{\bib}{\bibitem}
\nc{\al}{\alpha}
\nc{\g}{\gamma}
\nc{\G}{\Gamma}
\nc{\te}{\theta}
\nc{\D}{\Delta}
\nc{\eps}{\epsilon}
\nc{\la}{\lambda}
\nc{\La}{\Lambda}
\nc{\var}{\varphi}
\nc{\cg}{{\cal G}}
\nc{\cd}{{\cal D}}
\nc{\ep}{{\cal E}}
\nc{\pa}{\partial}
\nc{\nn}{\nonumber \\ }
\nc{\hf}{\frac{1}{2}}  
\nc{\dz}{\frac{dz}{2\pi i}}
\nc{\bin}[2]{\left (\begin{array}{c} {#1}\\ {#2} \end{array}\right )}
\nc{\ben}{\begin{equation}}
\nc{\een}{\end{equation}}
\nc{\bea}{\begin{eqnarray}}
\nc{\eea}{\end{eqnarray}}
\nc{\bra}[1]{\langle {#1}|}
\nc{\ket}[1]{|{#1}\rangle}
\newcommand{\Z}{\mathbb{Z}}
\nc{\C}{\mathbb{C}}
\nc{\Nat}{\mathbb{N}}

\nc{\HH}{\mathbb{H}}
\newcommand{\be}{\begin{eqnarray}}
\newcommand{\ee}{\end{eqnarray}}
\newcommand{\co}[2]{\lbrack #1 , #2 \rbrack}

\newcommand{\p}{\partial}
\def\vvdots{\mathinner{\mkern1mu\raise1pt\vbox{\kern7pt\hbox{.}}\mkern2mu
 \raise4pt\hbox{.}\mkern2mu\raise7pt\hbox{.}\mkern1mu}}

\begin{document}

\topmargin -5mm
\oddsidemargin 5mm

\begin{titlepage}
\setcounter{page}{0}

\vspace{8mm}
\begin{center}
{\huge On Stochastic Evolutions}\\[.4cm]
{\huge and Superconformal Field Theory}

\vspace{15mm}
{\Large Jasbir Nagi$^\ast$} and {\Large J{\o}rgen Rasmussen$^\dagger$}\\[.3cm] 
{\em $^\ast$DAMTP, University of Cambridge, Wilberforce Road}\\
{\em Cambridge, UK, CB3 0WA}\\[.3cm]
{\em $^\dagger$Centre de recherches math\'ematiques, Universit\'e 
 de Montr\'eal}\\ 
{\em Case postale 6128, 
succursale centre-ville, Montr\'eal, Qc, Canada H3C 3J7}\\[.3cm]
j.s.nagi@damtp.cam.ac.uk,\ \ \ rasmusse@crm.umontreal.ca

\end{center}

\vspace{10mm}
\centerline{{\bf{Abstract}}}
\vskip.4cm
\noindent
Links between certain stochastic evolutions of conformal maps
and conformal field theory have been studied in the realm of SLE
and by utilizing singular vectors in highest-weight modules
of the Virasoro algebra. It was recently found that this scenario 
could be extended to stochastic evolutions of 
superconformal maps of $N=1$ superspace with links to superconformal
field theory and singular vectors of the $N=1$ superconformal
algebra in the Neveu-Schwarz sector. Here we discuss the analogous
extension to the Ramond sector. We also discuss how the links are
modified when an unconventional superconformal structure or 
superderivative is employed.
\\[.5cm]
{\bf Keywords:} Stochastic evolutions, 
 superconformal field theory, superspace. 
\end{titlepage}
\newpage
\renewcommand{\thefootnote}{\arabic{footnote}}
\setcounter{footnote}{0}

\section{Introduction}

Certain two-dimensional systems at criticality may be described 
in terms of stochastic L\"owner evolutions (SLEs).
This was discovered by Schramm \cite{Sch} and has been developed
further in \cite{LSW,RS}, for example.
The method involves the study of stochastic evolutions of conformal maps.
A particular link to conformal field theory (CFT) has been examined
by Bauer and Bernard \cite{BB} (see also \cite{FW}) and extended
in \cite{LR}. In these studies, the SLE
differential equation is associated to a random walk on the 
Virasoro group. Based on the representation theory of CFT,
a more direct relationship may be established which involves
entities conserved in mean under the stochastic
process. A key ingredient is the existence of 
singular vectors in the associated Virasoro highest-weight modules.

This program was extended in \cite{Ras} to a link between 
stochastic evolutions of superconformal maps in $N=1$ 
superspace and the Neveu-Schwarz (NS) sector in 
$N=1$ superconformal field theory (SCFT).
This extension relies on an Ito calculus, not just for
commuting and anti-commuting variables as discussed in \cite{Rog}
and references therein, but for general non-commuting objects (e.g.
group- or algebra-valued entities) on superspace.
The generalization is straightforward, though, and discussed in \cite{Ras}.

Our main objective here is to work out the analogous results in
the Ramond sector of $N=1$ SCFT. 
This can be done by considering the superconformal condition in the
NS sector, but with the odd co-ordinate
on superspace having {\em anti-periodic} boundary conditions. Alternatively,
as was first noticed in \cite{KvL}, and developed further in \cite{Nag}, one can obtain
the Ramond sector by considering a different superconformal condition, in which case
the odd co-ordinate has {\em periodic} boundary conditions. 
The NS sector is then characterized by 
anti-periodic boundary conditions on the odd co-ordinate. In both approaches, we shall
use the representation theory of the $N=1$ superconformal
algebra in the Ramond sector 
to link stochastic evolutions in $N=1$ superspace to $N=1$ SCFT.
For completeness, we also address the similar link based on 
the alternative superconformal condition in the NS sector,
and review some of the original results of \cite{Ras}.

\section{SLE and CFT}

SLE \cite{Sch,LSW,RS}, or chordal SLE, describes
the evolution of boundaries of simply-connected domains
in the complex plane. 
This may be thought of as Brownian motion on the set of 
conformal maps $\{g_t\}$ satisfying the L\"owner equation
\ben
 \pa_tg_t(z)\ =\ \frac{2}{g_t(z)-\sqrt{\kappa}B_t},\ \ \ \ 
  \ \ \ \ \ g_0(z)\ =\ z
\label{g}
\een
$B_t$ is one-dimensional Brownian motion with $B_0=0$,
$\kappa\geq0$, and the system is often denoted SLE$_\kappa$.

The link to CFT \cite{BB} is established by considering
Ito differentials of Virasoro group elements $\cg_t$:
\ben
 \cg_t^{-1}d\cg_t\ =\ \left(-2L_{-2}+\frac{\kappa}{2}L_{-1}^2\right)dt
  +\sqrt{\kappa}L_{-1}dB_t,\ \ \ \ \ \ \ \cg_0=1
\label{G}
\een
Such group elements are obtained by exponentiating
generators of the Virasoro algebra 
\ben
 [L_n,L_m]\ =\ (n-m)L_{n+m}+ \frac{c}{12}n (n^2-1) \delta_{n+m,0}
\label{vir}
\een
where $c$ is the central charge.
The conformal transformation generated by a Virasoro group element
has a simple action on a primary field of weight $\Delta$.
For simplicity,
we do not distinguish explicitly between boundary and bulk primary
fields, nor do we write the anti-holomorphic part.
The transformation generated by $\cg_t$ reads
\ben
 \cg_t^{-1}\phi_\Delta (z)\cg_t\ 
  =\ \left(\partial_z f_t(z)\right)^\Delta \phi_\Delta (f_t(z))
\label{GphiG}
\een
for some conformal map $f_t$.
Using that the Virasoro generators act as
\ben
 [L_n,\phi_\Delta(z)]\ =\ (z^{n+1}\partial_z + \Delta(n+1)z^n) \phi_\Delta(z)
\label{Lphi}
\een
one finds that the conformal map $f_t$ associated to the random process 
$\cg_t$ (\ref{G}) must be a solution to the stochastic differential equation
\ben
 df_t(z)\ =\ \frac{2}{f_t(z)}dt-\sqrt{\kappa}dB_t,\ \ \ \ 
    \ \ \ \ \ f_0(z)\ =\ z
\label{df}
\een
With the definition
\ben
 f_t(z):=\ g_t(z)-\sqrt{\kappa}B_t
\label{f}
\een
this is merely a rewriting of the SLE equation (\ref{g}).

This link does not teach us to which CFT with central charge
$c$ a given SLE$_\kappa$ may be associated. 
A refinement is established by relating the representation theory of
the Virasoro algebra to entities conserved in mean under
the random process. Observables of the stochastic process (\ref{G}) 
are here thought of as functions on the Virasoro group.
With $\ket{\D}$ denoting the highest-weight 
vector of weight $\Delta$ in the Verma module ${\cal V}_\Delta$, 
we now follow \cite{BB} and consider the time evolution of the
expectation value of $\cg_t\ket{\D}$:
\ben
 \pa_t{\bf E}[\cg_t\ket{\Delta}]\ =\ {\bf E}[\cg_t\left(-2L_{-2}
  +\frac{\kappa}{2}L_{-1}^2\right)\ket{\Delta}]
\label{LD}
\een
The linear combination $-2L_{-2}+\frac{\kappa}{2}L_{-1}^2$ will
produce a singular vector when acting on the highest-weight vector
provided 
\ben
 c_\kappa\ =\ 1-\frac{3(4-\kappa)^2}{2\kappa},\ \ \ \ \ \ 
 \Delta_\kappa\ =\ \frac{6-\kappa}{2\kappa}
\label{cd}
\een
In this case the right-hand side of (\ref{LD}) vanishes
in the module obtained by factoring out
the singular vector from the reducible Verma module ${\cal V}_\Delta$.
The representation theory of the factor module is
thereby linked to the description of entities conserved
in mean.

\section{Superspace and SCFT}

Here we shall give a very brief summary of results 
from the theory of
$N=1$ superspace and $N=1$ SCFT required in
the following.
We refer to \cite{Dor,Nag} and references therein for 
recent accounts on the subject.
As we only consider $N=1$ superspace and $N=1$ SCFT
we shall simply refer to them as superspace and SCFT,
respectively.

\subsection{Superconformal maps}

Let 
\ben
 (z,\theta)\ \mapsto\ (z',\theta'),\ \ \ \ \ \ \ \ \ \left\{
   \begin{array}{ll} z'\ =\ g(z)+\theta\g(z)\\ \mbox{}\\
       \theta'\ =\ \tau(z)+\theta s(z) \end{array} \right.
\label{zt}
\een
denote a general superspace co-ordinate transformation. 
The conventionally associated superderivative reads
\ben
 D\ =\ \pa_\theta+\theta\pa_z
\label{D}
\een
Here $\theta$, $\theta'$, $\g$ and $\tau$ are anti-commuting
or (Grassmann) odd entities, while $z$, $z'$, $g$ and $s$ are even.
A superconformal transformation corresponding to (\ref{D})
is characterized by
\ben
 Dz'\ =\ \theta' D\theta'
\label{Dz'}
\een
in which case we have
\ben
 \g(z)\ =\ \tau(z)s(z),\ \ \ \ \ \ \pa_z g(z)\ =\ s^2(z)-\tau(z)\pa_z\tau(z)
\label{gts}
\een
and 
\ben
 D\ =\ (D\theta')D',\ \ \ \ \ \ D'\ =\ \pa_{\theta'}+\theta'\pa_{z'}
\label{D'}
\een
We shall be interested in locally invertible superconformal
maps implying that the complex part of $z'$ is non-vanishing
if the complex part of $z$ is non-vanishing.
We assume this is the case for $z\neq0$.

The superconformal transformations are generated
by $T$ and $G$ with modes $L_n$ and $G_r$
satisfying the superconformal algebra
\bea
 \left[L_n,L_m\right]&=&(n-m)L_{n+m}+ \frac{c}{12}n (n^2-1) \delta_{n+m,0}\nn
 \left[L_n,G_r\right]&=&(\frac{n}{2}-r)G_{n+r}\nn
 \left\{G_r,G_s\right\}&=&2L_{r+s}+\frac{c}{3}(r^2-\frac{1}{4})\delta_{r+s,0}
\label{LG}
\eea
The algebra is said to be in the Neveu-Schwarz (NS) sector when
the supercurrent $G$ has half-integer modes, and in the
Ramond sector when $G$ has integer modes. Branch cuts from square roots
are known to appear in the Ramond sector.

Below it will be discussed how one can avoid these branch 
cuts in the Ramond sector by introducing a different
superconformal structure. To this end, we consider the
superderivative
\ben
 {\cal D}\ =\ \p_{\theta} + \theta z\p_z
\label{Dalt}
\een
It is accompanied by a modified condition for a map (\ref{zt})
to be superconformal:
\ben\label{newscfcndn}
 {\cal D}z' \ =\  \theta'z'{\cal D}\theta '
\een
whereas (\ref{D'}) remains valid in the sense that 
\ben
 {\cal D}=({\cal D}\theta'){\cal D}',\ \ \ \ \ \ {\cal D}'\ =\ \pa_{\theta'}+\theta'z'\pa_{z'}
\label{cD}
\een
The generators of these superconformal transformations 
also satisfy (\ref{LG}). In particular, the modes of the supercurrent are
still integer in the Ramond sector and half-integer in the NS sector.
Branch cuts now appear in the NS sector instead of the Ramond sector,
and we have {\em periodic} 
boundary conditions on the odd co-ordinate in the Ramond sector,
and {\em anti-periodic} boundary conditions in the NS sector.
This is the opposite of the usual characterization of the sectors
based on the conventional superconformal structure defined by $D$.

\subsection{Primary fields and singular vectors}

The superconformal transformations generated
by Virasoro supergroup elements $\cg$ extend the ordinary
conformal case (\ref{GphiG}). 
Supergroup elements may be constructed by exponentiating
the generators of the superconformal algebra. 
Modes of the supercurrent must in that case be accompanied
by Grassmann odd parameters.
Primary fields are defined by having simple
transformation properties with respect to the superconformal
generators (\ref{LG}).
With $\D$ denoting the (super)conformal
dimension, a primary field corresponding
to the superderivative (\ref{D}) transforms as
\ben
 \cg^{-1}\Phi_\D(z,\theta)\cg\ 
  =\ \left(D\theta'\right)^{2\D} \Phi_\D(z',\theta')
\label{GPhiG}
\een
whereas it was found in \cite{Nag} that 
a primary field corresponding to the alternative
superderivative (\ref{Dalt}) transforms as 
\ben
 {\cal G}^{-1}\Phi_\D(z,\theta){\cal G}\ =\ 
   \left( \frac{z'}{z}({\cal D}\theta')^2\right)^\D\Phi_\D(z',\theta')
\label{ramprim}
\een
The action of the generators on a primary field thus depends
on the superderivative. From (\ref{GPhiG}) we have
\bea
 \left[L_n,\Phi_\D(z,\theta)\right]&=&\left(z^{n+1}\pa_z
   +\hf(n+1)z^n\theta\pa_\theta+\D(n+1)z^n\right)\Phi_\D(z,\theta)\nn
 \left[G_r,\Phi_\D(z,\theta)\right]&=&\left(
   -\theta z^{r+1/2}\pa_z+z^{r+1/2}\pa_\theta-\D(2r+1)\theta
   z^{r-1/2}\right)\Phi_\D(z,\theta)
\label{primary}
\eea
with $n\in\Z$ and $r\in\Z +\hf$,
while (\ref{ramprim}) corresponds to
\bea
 \co{L_n}{\Phi_\D(z,\theta)}&=&\left(z^{n+1}\p_z + \frac{n}{2}z^n\theta\p_{\theta}
  + \D(n+1)z^n\right)\Phi_\D(z,\theta)\nn
 \co{G_r}{\Phi_\D(z,\theta)}&=&\left(-\theta z^{r+1}\p_z + z^r\p_{\theta} -
  \D(2r+1)\theta z^r\right)\Phi_\D(z,\theta)
\label{supervirac}
\eea
where $n,r\in\Z$. The commutation relations of the operators
$L_n$, $G_r$ in (\ref{supervirac}) can be explicitly calculated and
hence can be shown to give rise
to a representation of the $N=1$ Ramond superconformal algebra.
Likewise, the more familiar
operators in (\ref{primary}) generate the $N=1$ NS superconformal algebra.
As indicated, we are considering {\em even} primary fields only.

The discussion above assumes periodic boundary conditions
on the odd co-ordinate and hence on the supercurrent
$G(z)=\sum_r G_r z^{-r-3/2}$, i.e., $G(ze^{2\pi i})=G(z)$.
For anti-periodic boundary conditions where $G(ze^{2\pi i})=-G(z)$,
the mode of $G_r$ in (\ref{primary}) would be integer (and thus
generate the Ramond algebra), while in (\ref{supervirac})
it would be half-integer (and thus generate the NS algebra).

A highest-weight module of the superconformal algebra is reducible
if it contains a submodule generated from a singular
vector. 
In the NS sector, the simplest non-trivial singular vector
in a highest-weight module with highest weight $\D$ 
appears at level 3/2 and is given by
\ben
 \ket{\chi;3/2}\ =\ \left((\D+\hf)G_{-3/2}-L_{-1}G_{-1/2}\right)\ket{\D}
\label{chi}
\een
provided
\ben
 (2\D+1)c\ =\ 3\D(3-2\D)
\label{Dc}
\een
In the Ramond sector, the similar vector appears at level 1:
\ben
 \ket{\chi;1}\ =\ \Big( (8\D+c)L_{-1} - 6G_{-1}G_0\Big) \ket{\D}
\label{chi1}
\een
provided 
\ben
 (16\D+3)c\ =\ 8\D(9-16\D)
\label{Dc1}
\een

\section{Graded stochastic evolutions and SCFT}

We first consider the stochastic differential
\ben
 \cg^{-1}_td\cg_t\ =\ \al dt+\sum_{i=1}^b\beta_i dB^{(i)}_t,\ \ \ \ \ \ \cg_0=1
\label{cg}
\een
which we can think of as a random walk on the Virasoro supergroup.
The coefficients $\al$ and $\beta_i$ are even and generically non-commutative
expressions in the generators of the superconformal algebra.
The differential is based on $b$-dimensional Brownian motion, 
$\bar{B}_t=(B^{(1)}_t,...,B^{(b)}_t)$, with $B_0^{(i)}=0$.
The associated Ito calculus treats the basic differentials according
to the rules
\ben
 (dt)^2\ =\ dtdB_t^{(i)}\ =\ 0,\ \ \ \ \ 
   dB_t^{(i)}dB_t^{(j)}\ =\ \delta_{ij}dt
\label{db}
\een
The Ito differential of the element inverse to $\cg_t$ is given by
\ben
 d(\cg^{-1}_t)\cg_t\ =\ 
  (-\al+\sum_{i=1}^b\beta_i^2)dt-\sum_{i=1}^b\beta_i dB^{(i)}_t
\label{cg-1}
\een

The superconformal transformation generated
by $\cg_t$ acts on a primary field as (\ref{GPhiG}) or (\ref{ramprim}),
which means that the new superspace co-ordinate 
$(z'_t,\theta'_t)$ becomes a
stochastic function of $(z,\theta)$, parameterized by
$t\in\mathbb{R}_{\geq 0}$. 
To relax the notation, the subscript $t$ will often be suppressed below.
A goal is to compute the Ito differential of
both sides of (\ref{GPhiG}) or (\ref{ramprim}) and thereby relate the
stochastic differential equations of $\cg_t$ and $(z',\theta')$.
In either case, the Ito differential of the left-hand side reads
\bea
 d\left(\cg^{-1}_t\Phi_\D\cg_t\right)&=&d(\cg^{-1}_t)\Phi_\D\cg_t+
  \cg^{-1}_t\Phi_\D d\cg_t+d(\cg^{-1}_t)\Phi_\D d\cg_t\nn
 &=&\left(-\left[\al_0,\cg^{-1}_t\Phi_\D\cg_t\right]
  +\hf\sum_{i=1}^b\left[\beta_i,\left[\beta_i,\cg^{-1}_t\Phi_\D\cg_t
    \right]\right]\right)dt\nn
   &&-\sum_{i=1}^b\left[\beta_i,\cg^{-1}_t\Phi_\D \cg_t\right]dB^{(i)}_t
\label{dleft}
\eea
where we have introduced $\al_0=\al-\hf\sum_{i=1}^b\beta_i^2$.
A direct comparison of (\ref{dleft}) with the Ito differential of the 
right-hand side of (\ref{GPhiG}) or (\ref{ramprim}) requires that
$\beta_i$ and $\al_0$ are linear in the generators:
\bea
 \al_0&=&\sum_{n}y_{0,n}L_n+\sum_r\eta_{0,r}G_{r}\nn
 \beta_i&=&\sum_{n}y_{i,n}L_n+\sum_r\eta_{i,r}G_{r}
\label{al0beta}
\eea
Here $n$ is integer while $r$ is integer or half-integer
depending on the sector.

With the subscript $t$ suppressed, we can express the
Ito differentials of the superspace co-ordinates as
\bea
 dz'&=&z_0'dt+\sum_{i=1}^bz_i'dB_t^{(i)},\ \ \ \ \ \ z'|_{t=0}\ =\ z\nn
 d\theta'&=&\theta_0'dt+\sum_{i=1}^b\theta_i'dB_t^{(i)},\ \ \ \ \ \ 
  \theta'|_{t=0}\ =\ \theta
\label{dzdt}
\eea
needed when computing the Ito differential of the 
right-hand sides of (\ref{GPhiG}) and (\ref{ramprim}).
The initial
conditions are required to match the initial condition on
$\cg$ in (\ref{cg}).

\subsection{General link for conventional superderivative $D$}

The computation of the Ito differential of the right-hand side of
(\ref{GPhiG}), corresponding to the conventional superderivative $D$,
was carried out in \cite{Ras}. The result for the NS sector
was subsequently compared
to (\ref{dleft}).
The calculation for the Ramond sector, where $r$ is integer rather
than half-integer, follows in exactly the same way. One uses the same
definition of primary field (\ref{primary}) and require that it be
single valued. The
same formulae are obtained, except replacing $r\in\Z+\hf$ with
$r\in\Z$. Without specifying this summation range explicitly,
the solutions in the two sectors may both be written as\footnote{Here and 
in the following, brackets may be used to write powers of $z'$ as $(z')^a$ 
instead of $z'^a$, for example.
The parameters preceding these expressions, such as
in $y_{i,n}(z')^{n+1}$, should thus not be confused with functions of $z'$.}
\bea
 z_i'&=&-\sum_{n\in\Z} y_{i,n}(z')^{n+1} + \sum_{r}\eta_{i,r}\theta' (z')^{r+\hf }\nn
 \theta_i'&=&-\hf\sum_{n\in\Z}(n+1)\theta'y_{i,n}(z')^n -
 \sum_{r}\eta_{i,r}(z')^{r+\hf}
\label{ziti}
\eea
and
\bea
 z_0'&=& -\sum_{n\in\Z} y_{0,n}(z')^{n+1} + \sum_{r}\eta_{0,r}\theta' (z')^{r+\hf }
  +\hf\sum_{i=1}^b\left(z_i'\pa_{z'}+\theta_i'\pa_{\theta'}
  \right)z_i'\nn
 \theta_0'&=&-\hf\sum_{n\in\Z} (n+1)\theta'y_{0,n}(z')^n -
 \sum_{r}\eta_{0,r}(z')^{r+\hf} 
  +\hf\sum_{i=1}^b\left(z_i'\pa_{z'}+\theta_i'\pa_{\theta'}
  \right)\theta_i'
\label{z0t0}
\eea
Using (\ref{ziti}), one may of course eliminate $z_i'$ and
$\theta_i'$ from these latter expressions, but the result is then less
compact than (\ref{z0t0}).
Only finitely many of the expansion coefficients $y$ and $\eta$
will be chosen non-vanishing rendering the sums in the solution 
(\ref{ziti}) and (\ref{z0t0}) finite.

In conclusion, the construction above establishes a general link between
a class of stochastic evolutions in superspace and SCFT.
The stochastic differentials (\ref{dzdt}) describing
the evolution of the superconformal maps are expressed
in terms of the parameters of the random walk
on the Virasoro supergroup (\ref{cg}), and this has
been achieved via the definition of primary
fields in SCFT (\ref{GPhiG}). The solution to the stochastic
differential equations in the NS sector is given in
(\ref{ziti}) and (\ref{z0t0}) when the sum over $r$
is over $r\in\Z+\hf$. The solution in the
Ramond sector, where the odd co-ordinate now
has anti-periodic boundary conditions, is obtained from the NS solution
by replacing the summation over $r\in\Z+\hf$ with $r\in\Z$.
As can be seen, this introduces square roots of $z'$, and
hence a branch cut in the Ramond solution.

\subsection{General link for alternative superderivative ${\cal D}$}

Rather than using (\ref{GPhiG}) and the conventional superderivative
$D$, one could try using (\ref{ramprim}) and the alternative superderivative 
$\mathcal{D}$. As we shall see, the branch cuts
in the Ramond sector encountered above are thereby avoided
but reappear in the NS sector instead.

To this end, we now consider the case where $(z', \te')$ and $\mathcal{G}$ 
in (\ref{ramprim}) evolve as stochastic processes, 
again parameterized by $t\in\mathbb{R}_{\geq 0}$. 
The Ito differential of the left-hand side of (\ref{ramprim})
is given by (\ref{dleft}), while the Ito
differential of the right-hand side is found to be
\be
 &&d\Big( \Big(\frac{z'}{z}\Big)^\D(\mathcal{D}\te ')^{2\D}\Phi(z',\theta')\Big) \nn
 &=& \frac{(\mathcal{D}\te')^{2\D}}{z^\D}\Big( dz'\{\D(z')^{\D-1}\Phi(z',\theta')
   + (z')^\D\p_{z'}\Phi(z',\theta')\} + d\te'\{(z')^\D\p_{\te'}\Phi(z',\theta')\} \nn
 &&+ (\mathcal{D}'d\te ')\{2\D(z')^\D\Phi(z',\theta')\} + dz'd\te'\{(z')^\D\p_{z'}\p_{\te'}
   \Phi(z',\theta') + \D(z')^{\D-1}\p_{\te'}\Phi(z',\theta')\}\nn
 &&+(dz')^2\{\frac{1}{2}\D(\D-1)(z')^{\D-2}\Phi(z',\theta') 
   + \frac{1}{2}(z')^\D\p_{z'}^2\Phi(z',\theta') +
  \D(z')^{\D-1}\p_{z'}\Phi(z',\theta')\}\nn
 && +(\mathcal{D}'d\te ')^2\{(z')^\D\D(2\D-1)\Phi(z',\theta')\} + (\mathcal{D}'d\te ')
   d\te'\{2\D(z')^\D\p_{\te'}\Phi(z',\theta')\}\nn
  && + dz'(\mathcal{D}'d\te ')\{2\D^2(z')^{\D-1}\Phi(z',\theta') 
    +2\D(z')^\D\p_{z'}\Phi(z',\theta')\}\Big)
\label{drightalt}
\ee
where we have used that $d(\cd\te')=\cd\te'\cd'(d\te')$ following from
(\ref{cD}).
Introducing the same expansions as in (\ref{dzdt}) and (\ref{al0beta}),
and using (\ref{supervirac}) and (\ref{db}),
one may now compare (\ref{dleft}) to (\ref{drightalt}). We thereby find
\be
 z'_i &=& -\sum_{n\in\Z} y_{i,n}(z')^{n+1} + \sum_r \eta_{i,r}\te'(z')^{r+1}\nn
 \te'_i &=& -\hf\sum_{n\in\Z} ny_{i,n}\te'(z')^n - \sum_r\eta_{i,r}(z')^r
\label{ziti2}
\ee
and
\be
 z'_0 &=& -\sum_{n\in\Z}y_{0,n}(z')^{n+1}+\sum_r\eta_{0,r}\te'(z')^{r+1}
   +\frac{1}{2}\sum_{i=1}^b\left(z'_i\p_{z'}
    + \te'_i\p_{\te'}\right)z'_i\nn
 \te'_0 &=& -\hf\sum_{n\in\Z}ny_{0,n}\te'(z')^n- \sum_r\eta_{0,r}(z')^r +
  \frac{1}{2}\sum_{i=1}^b\left(z'_i\p_{z'}+ \te'_i\p_{\te'}\right)\te'_i
\label{z0t02}
\ee
As in the case of the conventional superderivative above, 
$n$ is integer whereas $r$ is integer or half-integer
depending on the sector. The expressions (\ref{ziti2}) and
(\ref{z0t02}) thus represent the solution in the Ramond sector
when the sum over $r$ is over $\Z$, while $r\in\Z+\hf$ gives
the solution in the NS sector.
We wish to point out that, despite the use of a different superderivative
and a different definition of a primary field, there is a
striking similarity of these solutions with (\ref{ziti}) and
(\ref{z0t0}). However, the new results (\ref{ziti2}) and
(\ref{z0t02}) represent the Ramond sector
{\em without} the need of introducing a square root. 
Square roots appear in the NS sector instead.
This now gives an alternative
way of linking stochastic evolutions in superspace with SCFT.

\subsection{Expectation values and singular vectors}

Here we wish to indicate how time evolutions of expectation values 
may be evaluated. In this regard, observables of the process $\cg_t$ are
thought of as functions of $\cg_t$. The results in \cite{Ras}
on the NS sector, generalizing one of the main results in \cite{BB} on ordinary
SLE, may be extended straightforwardly to cover the Ramond sector
as well. We thus introduce the graded vector fields
\bea
 (\nabla_nF)(\cg_t)&=&\frac{d}{du}F(\cg_te^{uL_n})|_{u=0}\nn
 (\ep_rF)(\cg_t)&=&\frac{d}{d\nu}F(\cg_te^{\nu G_r})
   |_{\nu=0}
\label{nabla}
\eea
associated to the generators $L_n$ and $G_r$. Here
$\ep_r$ and $\nu$ (and of course $G_r$)
are odd, and $F$ is a
function admitting a `sufficiently convergent' Laurent expansion.
As always, the Ramond sector corresponds to $r$ being integer,
while half-integer $r$ corresponds to the NS sector.
Referring to the notation in (\ref{al0beta}), we find that
\ben
 \pa_t{\bf E}[F(\cg_t)]\ =\ {\bf E}[\left(\al_0(\nabla,\ep)+\hf\sum_{i=1}^b
   \beta_i^2(\nabla,\ep)\right)F(\cg_t)]
\label{nablaF}
\een
where
\bea
 \al_0(\nabla,\ep)&=&\sum_{n\in\Z}y_{0,n}\nabla_n+
  \sum_r \eta_{0,r}\ep_r\nn
 \beta_i(\nabla,\ep)&=&\sum_{n\in\Z}y_{i,n}\nabla_n+
   \sum_r\eta_{i,r}\ep_r
\label{alnabla}
\eea
This follows from (\ref{cg}), (\ref{cg-1}), 
$\al=\al_0+\hf\sum_{i=1}^b\beta_i^2$ and 
\bea
 \nabla_n(\cg^{-N}_t)&=&\frac{d}{du}
  \left(\left(e^{-uL_n}\cg_t^{-1}\right)^N\right)|_{u=0}\nn
  \ep_r(\cg^{-N}_t)&=&\frac{d}{d\nu}
  \left(\left(e^{-\nu G_r}\cg_t^{-1}\right)^N\right)|_{\nu=0}
\eea

This may be used to obtain a more direct relationship 
along the lines of section 2, thereby linking
the representation theory of the superconformal algebra,
through the construction of singular vectors, to entities
conserved in mean under the stochastic process.
In particular, it is seen that 
(\ref{nablaF}) reduces to
\ben
 \pa_t{\bf E}[\cg_t\ket{\Delta}]\ =\ {\bf E}[\cg_t\left(\al_0+
   \hf \sum_{i=1}^b\beta_i^2\right)\ket{\Delta}]
\label{EG}
\een
when $F(\cg_t)=\cg_t\ket{\D}$.
We should thus look for processes allowing us to put
\ben
 \left(\al_0+\hf \sum_{i=1}^b\beta_i^2\right)\ket{\D}\ \simeq \ 0
\label{0}
\een
in the representation theory. 
$\cg_t\ket{\D}$ is then a so-called martingale of the 
stochastic process $\cg_t$. That is, we should look for
processes where (\ref{0}) corresponds to a null vector
in the highest-weight module generated from the vector
 $\ket{\D}$.

\subsubsection{Results for conventional superderivative $D$}

It is discussed in \cite{Ras} how the NS singular
vector (\ref{chi}) may be obtained by considering
\ben 
 \al_0\ =\ -y\eta G_{-3/2},\ \ \ \ \ \ 
   \beta\ =\ \sqrt{\kappa}(yL_{-1}+\eta G_{-1/2}),\ \ \ \ \ \ y^2=0
\label{32}
\een
This corresponds to
\ben
 c_\kappa\ =\ \frac{15}{2}-3\left(\kappa+\frac{1}{\kappa}\right),
   \ \ \ \ \ \ \D_\kappa\ =\ \frac{2-\kappa}{2\kappa}
\label{cdk}
\een
and is in accordance with the ordinary labeling of reducible 
highest-weight modules of the superconformal algebra
with $\D=h_{3,1}$.
The associated stochastic differential equations are 
found in \cite{Ras} to be
\bea
 dz'&=&\frac{y\theta'\eta}{z'}dt-(y+\theta'\eta)\sqrt{\kappa}dB_t,\ \ \ \ \ 
  z'|_{t=0}\ =\ z\nn
 d\theta'&=&\frac{y\eta}{z'}dt-\eta\sqrt{\kappa}dB_t,\ \ \ \hspace{1.93cm}
  \theta'|_{t=0}\ =\ \theta
\label{d32}
\eea  
with solution
\bea
 z_t'&=&z+\frac{\theta y\eta}{z}t-(y+\theta\eta)\sqrt{\kappa}B_t\nn
 \theta'_t&=&\theta+\frac{y\eta}{z}t-\eta\sqrt{\kappa}B_t
\label{solNS}
\eea
Using $Dz'=\te'D\te'$, it is easily seen to be a superconformal map.

A different scenario based on a two-dimensional Brownian motion
is also discussed in \cite{Ras}. Here we shall confine ourselves to 
one-dimensional Brownian motion only, and 
our goal now is to examine the stochastic differential equations associated
to the level-one singular vector (\ref{chi1}) in the Ramond sector.
It may be constructed based on 
\ben
 \alpha_0\ =\ -\frac{1}{2}\epsilon\eta L_{-1}\ ,\ \ \qquad \beta\ =\
  \sqrt{\kappa}(\epsilon G_{-1} + \eta G_0)
\label{ab1}
\een
as one then has
\ben
 \Big( \alpha_0 + \frac{1}{2}\beta^2 \Big)\ket{\D}\ =\
  \epsilon\eta\Big((\kappa-\hf)L_{-1}-\kappa G_{-1}G_0 \Big)\ket{\D}
\label{absing}
\een
This corresponds to
\ben
 c_\kappa\ =\ \frac{15}{2}-3\left(\kappa+\frac{1}{\kappa}\right),     
    \ \ \ \ \ \ \D_\kappa\ =\ \frac{6\kappa-3}{16}
\label{cdk1}
\een
and is in accordance with the ordinary labeling of reducible 
highest-weight modules of the superconformal algebra
with $\D=h_{1,2}$.
The associated stochastic differential equations are 
found to be
\bea 
 dz'&=& \frac{1}{2}\epsilon\eta dt + \sqrt{\kappa}\left(\epsilon
   (z')^{-\frac{1}{2}} + \eta (z')^{\frac{1}{2}}\right)\te'dB_t,\ \ \ \hspace{0.96cm}
  z'|_{t=0}\ =\ z\nn
 d\te '&=&\frac{\kappa\epsilon\eta\te'}{2z'} dt
  - \sqrt{\kappa}\left( \epsilon (z')^{-\frac{1}{2}} + \eta
  (z')^{\frac{1}{2}}\right) dB_t, \ \hspace{1.2cm}
  \theta'|_{t=0}\ =\ \theta
\label{infchange}
\eea
Substituting
\ben
 w'\ =\ z' - \sqrt{\kappa}\left( \epsilon (z')^{-\frac{1}{2}} +
\eta (z')^{\frac{1}{2}} \right)\te' B_t
\een
it can be shown that
\ben
 dw' \ =\ \frac{1}{2}\epsilon\eta dt,\ \ \ \ \ \ \ \ 
  w'|_{t=0}\ =\ z
\een
which can be readily solved:
\ben
 w'\ =\ z + \frac{1}{2}\epsilon\eta t
\label{wsoln}
\een
Making the additional substitution
\ben
 \chi' \ =\ \theta' + \sqrt{\kappa}\left( \epsilon (z')^{-\frac{1}{2}} + \eta
  (z')^{\frac{1}{2}}\right) B_t
\label{chidef}
\een
we have 
\ben
 d\chi'\ =\ -\frac{\kappa\epsilon\eta\te' }{2z'}  dt - 
    \frac{\kappa\epsilon\eta\te'B_t }{z'}  dB_t,
     \ \ \ \ \ \ \  \chi'|_{t=0}\ =\ \theta
\label{chieqn}
\een
with solution
\ben
 \chi'\ =\ \te - \frac{\kappa\epsilon\eta\te' B_t^2}{2z'}
\label{chisoln}
\een
The original stochastic differential equations (\ref{infchange}) are finally
found to be solved by
\bea
 z' &=& z + \frac{1}{2}\epsilon\eta t + \sqrt{\kappa} \left(\epsilon
  z^{-\frac{1}{2}} + \eta z^{\frac{1}{2}}\right)\te B_t\nn
 \te '&=& \te - \sqrt{\kappa}\left(\epsilon z^{-\frac{1}{2}} + \eta
  z^{\frac{1}{2}}\right)B_t + \frac{\kappa\epsilon\eta\te}{2z}B_t^2
\label{soln1}
\eea
This is easily seen to correspond to a superconformal map.
It is observed that, unlike the NS solution (\ref{solNS}), 
this Ramond solution involves the Brownian motion {\em squared}.
It is also noted that the solution represents stochastic evolutions 
in only the `soul' part of the even co-ordinate, as the complex
part of it remains unchanged. As already announced, square
roots of $z$ are apparent in these expressions.

\subsubsection{Results for alternative superderivative $\cd$}

We now turn to the alternative superderivative $\cd$ using
(\ref{ziti2}) and (\ref{z0t02}). First we consider the Ramond sector
based on the level-one singular vector (\ref{chi1}) and the random
walk given by (\ref{ab1}) and (\ref{absing}). The associated
stochastic differential equations read
\bea
 dz' &=& \frac{1}{2}\epsilon\eta dt + \sqrt{\kappa}(\eta z' + \epsilon)\te'dB_t,
  \ \ \ \ \ \ \ \ \ \ \ \ \hspace{1.07 cm} z'|_{t=0}\ =\ z\nn
 d\te'&=&\left( \kappa -\frac{1}{2}\right) \frac{\epsilon\eta\te '}{2z'}dt
   - \sqrt{\kappa}\left(\eta + \frac{\epsilon}{z'}\right)dB_t,
      \ \ \ \ \ \ \  \te'|_{t=0}\ =\ \theta
\label{eqn1}
\eea
which are seen to be free of square roots of $z'$, as expected.
These equations can be solved in a manner similar to the one employed above.
First we substitute
\ben
 w' \ =\ z' - \sqrt{\kappa}(\eta z' + \epsilon)\te 'B_t
\een
yielding the stochastic equation
\ben
 dw'\ =\ \frac{1}{2}\epsilon\eta dt ,\ \ \ \ \ \ \ \ \
  w'|_{t=0}\ =\ z
\een
with solution
\ben
 w'\ =\ z +\hf\epsilon\eta t
\een
Making the further substitution
\ben
 \chi '\ =\ \te ' + \sqrt{\kappa}\left(\eta + \frac{\epsilon}{z'}\right)B_t +
\frac{\kappa\epsilon\eta\te '}{2z'}B_t^2 
\een
yields the stochastic equation
\ben
 d\chi'\ =\ -\frac{\epsilon\eta\te'}{4z'}dt,\ \ \ \ \ \ \ \ \  \chi'|_{t=0}\ =\ \theta
\een
with solution
\ben 
 \chi'\ =\ \te-\frac{\epsilon\eta\te't}{4z'}
\een
The solution to the original set of equations (\ref{eqn1}) finally reads
\bea
 z'&=& z + \frac{1}{2}\epsilon\eta t + \sqrt{\kappa}\left(\eta z +
   \epsilon\right)\te B_t\nn
 \te'&=& \te-\frac{\epsilon\eta\te}{4z}t - \sqrt{\kappa}\left(\eta +
   \frac{\epsilon}{z}\right)B_t + \frac{\kappa\epsilon\eta\te}{2z}B_t^2
\label{soln3}
\eea
This can be substituted directly into
(\ref{newscfcndn}) to check that it corresponds to a superconformal map. 
Two different,
stochastically evolving systems, (\ref{soln1}) and (\ref{soln3}), 
have thus been found to correspond to the {\em same}
singular vector (\ref{chi1}). Only one of them contains square roots of $z$. 

To `complete the picture', we now examine the
stochastic differential equations following from considering
the NS singular vector (\ref{chi}) in the realm of 
the alternative superconformal structure based on $\cd$.
Following our general prescription, we find
\bea
 dz'&=&-y\eta\te'(z')^{-\hf}dt+\sqrt{\kappa}\left(\eta\te'(z')^{\hf}-y\right)dB_t,
   \ \ \ \ \ \hspace{2.67cm} z'|_{t=0}\ =\ z\nn
 d\te'&=&(1-\frac{\kappa}{2})y\eta(z')^{-\frac{3}{2}}dt
  +\sqrt{\kappa}\left(\hf y\te'(z')^{-1}-\eta(z')^{-\hf}\right)dB_t,
      \ \ \ \ \ \ \  \te'|_{t=0}\ =\ \theta
\label{dzNSalt}
\eea
The new co-ordinate
\ben
 w'\ =\ z'+\sqrt{\kappa}\left(y-\eta\te'(z')^{\hf}\right)B_t
\een
then satisfies the stochastic differential equation
\ben
 dw'\ =\ -y\eta\te'(z')^{-\hf}dt,  \ \ \ \ \ \ \ \ \ w'|_{t=0}\ =\ z
\een
with solution
\ben
 w'\ =\ z-y\eta\te'(z')^{-\hf}t
\een
Likewise, the substitution
\ben
 \chi'\ =\ \te'+\sqrt{\kappa}\left(\eta(z')^{-\hf}-\hf y\te'(z')^{-1}\right)B_t,
   \ \ \ \ \ \ \ \ \ \chi'|_{t=0}\ =\ \te
\een
must respect
\ben
 d\chi'\ =\ (1+\frac{\kappa}{2})y\eta(z')^{-\frac{3}{2}}dt
   +\kappa y\eta(z')^{-\frac{3}{2}}B_tdB_t
\een
with solution
\ben
 \chi'\ =\ \te+y\eta(z')^{-\frac{3}{2}}t+\hf\kappa y\eta(z')^{-\frac{3}{2}}B^2_t
\een
One could have made a further substitution and considered the equation
\ben
 d\left((w')^{\frac{3}{2}}\tilde{\chi}'\right)\ =\ 
  d\left((w')^{\frac{3}{2}}\{\chi'-\frac{\kappa}{2}y\eta(z')^{-\frac{3}{2}}B_t^2\}\right)\ =\
  y\eta dt,\ \ \ \ \ \ \  (w')^{\frac{3}{2}}\tilde{\chi}'|_{t=0}\ =\ z^{\frac{3}{2}}\te
\een
which is solved trivially and without reference to the original equations (\ref{dzNSalt}).
In either case, one may conclude that the original equations are solved by
\bea
 z'&=&z-y\eta\te z^{-\hf}t+\sqrt{\kappa}\left(\eta\te z^\hf-y\right)B_t\nn
 \te'&=&\te+y\eta z^{-\frac{3}{2}}t+\sqrt{\kappa}\left(\hf y\te z^{-1}-\eta z^{-\hf}\right)B_t
  -\hf\kappa y\eta z^{-\frac{3}{2}}B_t^2
\eea
This is seen to correspond to a superconformal map.

\section{Conclusion}

The results of \cite{Ras} on the links between stochastic evolutions of superconformal
maps and the NS sector of $N=1$ SCFT have been extended 
to the Ramond sector, as well as to an unconventional superconformal structure
underlying $N=1$ superspace and $N=1$ SCFT. 
In the usual treatment of the Ramond sector,
it is found that square roots of
the even co-ordinate, $z$, have to be introduced. 
Based on the alternative superconformal structure and using 
the associated definition of a primary field found in
\cite{Nag}, a modified set of equations are found.
They give results for the Ramond sector {\em without} square roots,
while square roots appear in the NS sector. 
To the best of our knowledge,
this is the first time that such a primary field has been used, and
results compared to the more conventional treatment of the Ramond 
and NS sectors.
\vskip.5cm
\noindent{\em Acknowledgements}
\vskip.1cm
\noindent JN is funded by PPARC and thanks them for their financial support.
The authors thank the EUCLID network, contract HPRN-CT-2002-00325,
for funding the first EUCLID network conference (Firenze, 2003) where the present collaboration was conceived.

\end{document}